\documentclass[conference]{IEEEtran}

\usepackage[pdftex]{graphicx}
\usepackage{xcolor}
\usepackage[hidelinks]{hyperref}

\usepackage{balance}

\definecolor{Paired-1}{RGB}{31,120,180}
\definecolor{Paired-2}{RGB}{166,206,227}
\definecolor{Paired-3}{RGB}{51,160,44}
\definecolor{Paired-4}{RGB}{178,223,138}
\definecolor{Paired-5}{RGB}{227,26,28}
\definecolor{Paired-6}{RGB}{251,154,153}
\definecolor{Paired-7}{RGB}{255,127,0}
\definecolor{Paired-8}{RGB}{253,191,111}
\definecolor{Paired-9}{RGB}{106,61,154}
\definecolor{Paired-10}{RGB}{202,178,214}
\definecolor{Paired-11}{RGB}{177,89,40}
\definecolor{Paired-12}{RGB}{255,255,153}

\usepackage{amsmath,amssymb,amsfonts,stmaryrd}
\usepackage{bm}
\usepackage{dsfont}
\usepackage{numprint}
\newcommand{\isep}{\mathrel{{.}\,{.}}\nobreak}

\usepackage{adjustbox}
\usepackage{booktabs}
\usepackage{multirow}

\ifCLASSOPTIONcompsoc
    \usepackage[caption=false,font=normalsize,labelfont=sf,textfont=sf]{subfig}
\else
    \usepackage[caption=false,font=footnotesize]{subfig}
\fi

\usepackage{pgfplots}
\pgfplotsset{compat=newest}
\usepgfplotslibrary{groupplots}

\usepackage{tikz}
\usetikzlibrary{matrix, positioning, patterns, shapes, arrows}

\usepackage[draft]{fixme} 
\fxsetup{theme=color}

\usepackage{lipsum}

\usepackage[ruled, linesnumbered, vlined]{algorithm2e}

\title{High-Throughput VLSI Architecture for GRAND}

\author{\IEEEauthorblockN{Syed Mohsin Abbas, Thibaud Tonnellier, Furkan Ercan, and Warren J. Gross}
\IEEEauthorblockA{Department of Electrical and Computer Engineering\\
McGill University, Montréal, Québec, Canada\\
Emails: syed.abbas@mail.mcgill.ca, thibaud.tonnellier@mcgill.ca, furkan.ercan@mail.mcgill.ca, warren.gross@mcgill.ca}}

\begin{document}

\maketitle

\begin{abstract}
Guessing Random Additive Noise Decoding (GRAND) is a recently proposed universal decoding algorithm for 
linear error correcting codes. 
Since GRAND does not depend on the structure of the code, it can be used for any code encountered in contemporary
communication standards or may even be used for random linear network coding. This property makes this new algorithm particularly appealing.
Instead of trying to decode the received vector, GRAND attempts to identify the noise that corrupted the codeword.
To that end, GRAND relies on the generation of test error patterns that are successively applied to the received vector.
In this paper, we propose the first hardware architecture for the GRAND algorithm. 
Considering GRAND with ABandonment (GRANDAB) that limits the number of test patterns, the proposed architecture only needs 
$2+\sum_{i=2}^{n} \left\lfloor\frac{i}{2}\right\rfloor$ time steps to perform the $\sum_{i=1}^3 \binom{n}{i}$ queries 
required when $\text{AB}=3$. For a code length of $128$, our proposed hardware architecture demonstrates only a fraction 
($1.2\%$) of the total number of performed queries as time steps. 
Synthesis result using TSMC 65nm CMOS technology shows that average throughputs of $32$ Gbps to $64$ Gbps 
can be achieved at an SNR of $10$ dB for a code length of $128$ and code rates rate higher than $0.75$, transmitted over an AWGN channel.
Comparisons with a decoder tailored for a $(79,64)$ BCH code show that the proposed architecture can achieve a slightly higher 
average throughput at high SNRs, while obtaining the same decoding performance. 
\end{abstract}


\section{Introduction}
Since the landmark paper by Shannon \cite{Shannon48} in 1948, one of the goal of researchers in the field of information theory 
was to find good error correcting codes that can be efficiently decoded. As soon as 1950, Hamming proposed his eponymous codes that 
can always correct one error \cite{Hamming50}. Ten years later,  Bose–Chaudhuri–Hocquenghem (BCH) codes 
\cite{Hocquenghem59,Bose1960} were discovered. They have the pleasant property that the number of correctable errors is 
chosen by design. To find the locations of the errors, two main algorithms can be considered: the Berlekamp–Massey 
algorithm \cite{Berlekamp68,Massey69} or the Peterson-Gorenstein–Zierler (PGZ) algorithm \cite{Peterson60}. 
After major advances and rediscoveries in the 90's, polar codes were proposed in 2008
along with their decoding algorithm called successive cancellation (SC) algorithm \cite{Arikan09}. This is the first proven 
class of codes that asymptotically reaches the Shannon limit. Serially concatenated with an outer cyclic redundancy check
(CRC) code \cite{Peterson61}, polar codes have been selected as part of the 5G New Radio (NR) standard \cite{3GPP}.
All of these algorithms require devoted decoding techniques, and a decoder tailored for a decoding algorithm cannot be 
directly utilized for another one. 

Recently, a universal decoding algorithm for linear codes has been proposed \cite{Duffy19TIT}. Named Guessing Random 
Additive Noise Decoding (GRAND), the algorithm does not rely on the underlying channel code. Instead of using the 
properties and the structure of the code to identify the errors that may occur at the reception due to the channel 
noise, GRAND guesses the noise present in the received vector. In other words, GRAND is noise-centric rather
than being code-centric and is able to tackle any aforementioned coding schemes that have been developed over the course 
of information theory. GRAND is not the only decoding algorithm that is code agnostic. 
However, considering high rate codes, GRAND has a lower computational complexity than a brute-force search \cite{Duffy19TIT},
or does not require the costly Gaussian elimination required for information set decoding \cite{Prange62}. 
The Gaussian elimination is also required for random linear network coding \cite{Ho03} and GRAND could be an efficient 
way to reduce the computational complexity of this powerful encoding scheme. 
In addition, an approach similar to GRAND has been used to lower the error floor of turbo codes in \cite{Tonnellier16}, which could lead to the use of GRAND in conjunction with usual decoding techniques.

To identify the noise, GRAND has three main steps. First, error patterns are generated in a specific order. Then, the 
error patterns are combined with the received vector, and finally, queries for codebook membership on the resulting words 
are realized. Generating all the possible error patterns is impractical and unwanted. Thus, GRAND with ABandonment 
(GRANDAB) has been also proposed to limit the number of queries performed during the process \cite{Duffy19TIT}.

In \cite{Duffy20205g}, the application of GRANDAB for short length and high rate CRC-polar codes encountered in the 5G 
NR standard has been demonstrated. Considering a code of length $128$ and up to $3$ errors, a maximum number of $\numprint{349632}$ queries
may be required. However, it has been observed that the average number of queries are much smaller than the worst-case 
scenario for practical signal-to-noise ratio (SNR) conditions. Thus, GRANDAB offers a high throughput with an average low 
latency at moderate-to-high SNR regimes, and tailored for high code rates, both of which are particularly crucial for 
ECC storage applications.

In \cite{duffy2020ordered} and in \cite{solomon2020soft}, GRAND is enhanced to consider soft-information at its 
input. Impressive decoding performance is achieved, where substantial gains over the SC-List \cite{Tal15} are presented. However, 
we limit the scope of this work to hard input decoding.

In this paper, we propose a high throughput hardware architecture for GRANDAB. To this end, we first show how to share 
computations required by the GRAND algorithm, using basic linear algebra. Then we propose an efficient hardware exploiting
the proposed sharings. To the best of our knowledge, this is the first hardware architecture implementing the GRAND algorithm.
Considering a code of length $128$, and with a correction capability of up to three errors, the proposed architecture can
achieve an average coded throughput of up to 64 Gbps. Moreover, the proposed architecture can achieve the same average throughput
as a recently proposed decoder that can only consider a $(79,64)$ BCH code.

The rest of this work is organized as follows: In Section II, preliminaries regarding linear codes and GRAND algorithms 
are given. In Section III, the proposed hardware architecture is detailed. Synthesis results and comparison with a
state-of-the-art decoder for BCH code are given.
Finally, concluding remarks are drawn in Section V.

\section{Preliminaries}
\subsection{Notations}
Matrices are denoted by a bold upper-case letter ($\bm{M}$), while vectors are denoted with bold lower-case letters ($\bm{v}$).
The transpose operator is represented by $^\top$. 
The number of $k$-combinations from a given set of $n$ elements is noted by $\binom{n}{k}$.
$\mathds{1}_n$ is the indicator vector where all locations except the $n^{\text{th}}$ are $0$ and the the $n^{\text{th}}$ is $1$. 
All the indices start at $1$.

\subsection{Linear block codes}
Due to their convenient representations, linear block codes are a class of error-correcting codes widely adopted by communication standards. 
In the following, we restrict ourselves 
with operations in the Galois field with 2 elements, noted $\mathbb{F}_2$. A block code is a mapping $g: \mathbb{F}_2^k \rightarrow \mathbb{F}_2^n$, 
where $k < n$. This way, a vector $\bm{u}$ of size $k$ maps to a vector $\bm{c}$ of size $n$.
The set of the $2^k$ vectors $\bm{c}$ is called a code $\mathcal{C}$, whose elements $\bm{c}$ are called
 codewords. The ratio $R \triangleq \frac{k}{n}$ is the code rate. 
 If $g$ is a linear mapping, then $\mathcal{C}$ is a linear block code. Thus, there exists a $k \times n$ matrix $\bm{G}$
called generator matrix of the code $\mathcal{C}$. Then, the encoding process can be realized as a vector-matrix product: $\bm{c} = \bm{u}\cdot\bm{G}$.
We can define $\bm{H}$, the $(n-k) \times n$ generator matrix of the dual code of $\mathcal{C}$. $\bm{H}$ is also called the parity-check matrix 
of $\mathcal{C}$ and verifies the following property:
\begin{equation}
\forall~\bm{c} \in \mathcal{C},~\bm{H}\cdot\bm{c}^\top = \bm{0}.
\label{eq:pcheck}
\end{equation}

Consider that $\bm{c}$ has been transmitted over a noisy channel and that $\bm{r}$ is received at the output 
of the channel. Because of the channel noise, $\bm{r}$ can differ from $\bm{c}$. Therefore, we can establish the relationship 
between $\bm{r}$ and $\bm{c}$ as: $\bm{r} = \bm{c}~\oplus~\bm{e}$, where
$\bm{e}$ is the \emph{error pattern} caused by the channel noise. The \emph{syndrome} is defined by $\bm{s} \triangleq \bm{H}\cdot\bm{r}^\top$.
According to (\ref{eq:pcheck}), $\bm{s}$ is zero if and only if $\bm{r}$ is a codeword. Thus, if $\bm{s}$ is zero
either there is no error or the error pattern is itself a codeword. This is the basic principle for standard array decoding
\cite{Slepian56}, and also for GRAND.

\subsection{Maximum Likelihood Decoding via GRAND}
Guessing Random Additive Noise Decoding (GRAND) is a recently proposed hard detection decoder that has been proven to 
be a maximum likelihood (ML) decoder \cite{Duffy19TIT}. Algorithm \ref{alg:grand} summarizes the steps of 
the GRAND procedure. The 
principle of GRAND is to generate test error patterns, to apply them to the received vector, and to check if the 
generated candidate is a codeword by verifying that
\begin{equation}
\bm{H} \cdot(\bm{r} \oplus \bm{e})^\top
\label{eq:constraint}
\end{equation}
is equal to zero. If so, $\hat{\bm{c}} \triangleq \bm{r}~\oplus~\bm{e}$ is the estimated codeword. To perform a proper decoding, 
$\hat{\bm{c}}$ has to be converted into 
the estimated message: $\hat{\bm{u}} \triangleq \hat{\bm{c}}\cdot\bm{G}^{-1}$, where $\bm{G}^{-1}$ is the $n\times k$ matrix such that
$\bm{G}\cdot\bm{G}^{-1}$ is the identity matrix of size $k$. Note that this step is not required for systematic codes,
since the message bits directly appear in the codeword. 

\begin{algorithm}[t]
\caption{\label{alg:grand}GRAND for linear codes}
    \DontPrintSemicolon
    \SetAlgoVlined  
    \SetKwData{e}{$\bm{e}$}
    \SetKwData{estm}{$\hat{\bm{u}}$}
    \SetKwData{ginv}{$\bm{G}^{-1}$}
    \KwIn{$\bm{H}$, \ginv, $\bm{r}$}
    \KwOut{\estm}
    \SetKwFunction{RecursiveComputeLLRs}{recursiveComputeLLRs}
    \SetKwFunction{DecodeRONE}{decodeR1}
    \SetKwFunction{RDecodeRONE}{redecodeR1}
    \SetKwFunction{DecodeRZERO}{decodeR0}
    \SetKwFunction{Find}{findCandidate}
    \SetKwFunction{new}{generateNewErrorPattern}
    $\e \leftarrow \bm{0}$\;
    \While{$\bm{H} \cdot(\bm{r} \oplus \e)^\top \neq \bm{0}$}{
        $\e \leftarrow$ \new{}\;
    }
    $\estm \leftarrow (\bm{r} \oplus \e)\cdot\ginv$\;
    \KwRet{\estm}
\end{algorithm}

The most important property of GRAND is that it requires no other condition on the code except the linearity. Thus, GRAND can be 
considered for any linear code if only the parity check matrix ($\bm{H}$) is provided. To the best of our knowledge, the only
other decoders that can decode any linear code are the brute-force ML decoder and the information 
set decoding \cite{Prange62}, or its more recent version known as ordered statistic decoding \cite{Fossorier95}.
However, the ML decoder is impractical for high-rates codes since the $2^k$ codewords have to be evaluated, while the 
two others require Gaussian elimination, which is challenging to efficiently implement in hardware \cite{Scholl13}.

\subsection{GRAND with Abandonment}\label{sec:grandab}
To limit the computational complexity of the GRAND algorithm, GRAND with ABandonment (GRANDAB) is also proposed in \cite{Duffy19TIT}. 
In that case, the decoder abandons the search for the error pattern after a fixed number of queries is reached. Therefore, 
GRANDAB results in an approximated ML decoding. The notation GRANDAB with $\text{AB} = t$, means that 
the Hamming weight of the considered error patterns do not exceed $t$. 
As a result, the maximum number of queries for a code of length $n$ is given by
\begin{equation}
\sum\limits_{i=1}^t \binom{n}{i}.
\label{eq:queries} 
\end{equation}

\begin{figure}[!t]
\begin{tikzpicture}
    \begin{semilogyaxis}[
            footnotesize, width=\columnwidth, height=.7\columnwidth,    
            xmin=4, xmax=12.5, xtick={4,5,...,12},
            xlabel=SNR (dB), ylabel=FER,  
            grid=both, grid style={gray!30},
            tick align=outside, tickpos=left, 
            legend pos=south west, legend cell align={left},
            /pgfplots/table/ignore chars={|},
        ]

        \addplot[mark=diamond , Paired-7, semithick]  table[x=SNR, y=FER] {data/crc/128_120_SNR.txt};
        \addplot[mark=triangle, Paired-5, semithick]  table[x=SNR, y=FER] {data/crc/128_112_SNR.txt};
        \addplot[mark=square  , Paired-3, semithick]  table[x=SNR, y=FER] {data/crc/128_104_SNR.txt};
        \addplot[mark=o       , Paired-1, semithick]  table[x=SNR, y=FER] {data/crc/128_96_SNR.txt};

         \legend{{} {CRC(128,120)},{} {CRC(128,112)},{} {CRC(128,104)},{} {CRC(128,\hphantom{1}96)}}
    \end{semilogyaxis}
\end{tikzpicture}  
\caption{\label{fig:fer_crc}Comparison of the GRANDAB ($\text{AB}=3$) decoding performance using CRC codes for several rates and N=128.}
\end{figure}
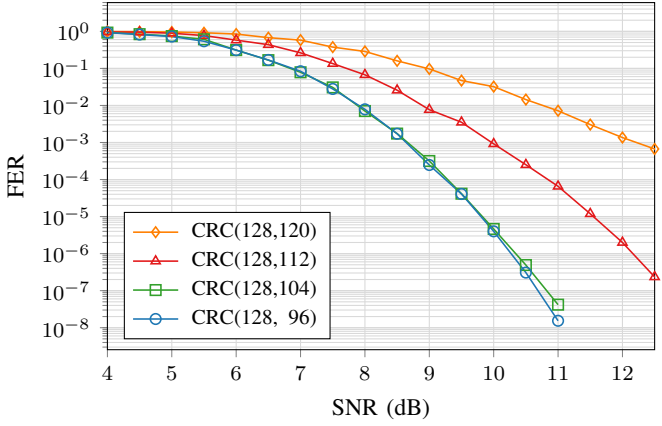

For an illustrative purpose, Fig. \ref{fig:fer_crc} compares the frame error rate (FER) performance obtained with 
GRANDAB $\text{AB} = 3$, using CRC codes with different rates ($R \geq 0.75)$. 
All codes have a length $n$ of 128. The generator polynomials are \texttt{0x04C11DB7}, \texttt{0xB2B117},
\texttt{0x1021}, and \texttt{0xD5} for $k=96$, $k=104$, $k=112$, and $k=120$, respectively.
A BPSK modulation and an AWGN channel with variance $\sigma^2$ are considered. The SNR in dB is defined as 
$\text{SNR} = -10\log_{10}\sigma^2$. The demodulator provides hard decisions to the GRANDAB decoder. 
Observe that the FER performance improves with the number of redundancy bits, up to the case $\text{CRC}(128, 104)$. 
Considering more redundancy bits does not improve the decoding performance since the considered version of 
GRANDAB cannot correct more than 3 errors. Nevertheless, GRANDAB is an effective way for decoding any code, especially
compared to its high-complexity agnostic hard decoder counterparts, for which an impractical number of computations are 
required. With $n=128$ and $\text{AB}=3$, a total number of 
$\numprint{349632}$ queries are required. Despite this large number of queries, targeting a FER of $10^{-4}$, 
the \textit{average} number of queries become $445$, $412$, $4.58$, and $1.01$ for $k=96$, $k=104$, $k=112$, and $k=120$, 
respectively. This is another advantage of GRAND: the average computational complexity decreases sharply as channel conditions improve.

\section{VLSI Architecture for GRAND}
In this section, we provide details of the proposed VLSI architecture for GRANDAB ($\text{AB} = 3$) decoding of linear codes.
Since GRAND decoding is agnostic to the underlying channel code, the proposed VLSI architecture can be used to decode 
any linear block code conforming with the length and rate constraints, given the parity check matrix $(\bm{H})$ of that code. 
Before presenting the details of our proposed VLSI architecture, we provide a minimal mathematical background 
-- exploiting the linearity of the considered codes -- required to simplify the problem.



\subsection{Computations reformulation}
For the one bit-flip error patterns $\mathds{1}_i$, with $i \in \llbracket 1\isep n \rrbracket$, using the distributivity rule,
(\ref{eq:constraint}) can be written as
\begin{equation}
\bm{H} \cdot(\bm{r} \oplus \mathds{1}_i)^\top=\bm{H}\cdot\bm{r}^\top \oplus \bm{H}\cdot\mathds{1}_i^\top, 
\label{eq:one-bit-constraint}
\end{equation}
where $\bm{H}\cdot\bm{r}^\top$ is the $(n-k)$-bits syndrome associated with the received
vector $\boldsymbol{r}$ and $\bm{H}\cdot\mathds{1}_i^\top$ is the
$(n-k)$-bits syndrome associated with the one bit-flip error pattern $\mathds{1}_i$.

Noticing that the two bit-flips noise sequences $\mathds{1}_{i,j}$, with $i \in \llbracket 1\isep n \rrbracket$, 
$j \in \llbracket 1\isep n \rrbracket$ and $i\neq j$, can be written as $\mathds{1}_{i,j} = \mathds{1}_i \oplus \mathds{1}_j$,
(\ref{eq:constraint}) can be expressed as 
\begin{equation}
\bm{H} \cdot(\bm{r} \oplus \mathds{1}_{i,j})^\top=\bm{H}\cdot\bm{r}^\top \oplus \bm{H}\cdot\mathds{1}_i^\top \oplus \bm{H}\cdot\mathds{1}_j^\top,
\label{eq:two-bits-constraint}
\end{equation}
for the two bit-flips case. Similarly, the three-bit-flips noise sequences $\mathds{1}_{i,j,k}$,
where $i$, $j$ and $k$ are the flipped bit positions, can be checked
for code membership with 
\begin{equation}
\bm{H} \cdot(\bm{r} \oplus \mathds{1}_{i,j,k})^\top=\bm{H}\cdot\bm{r}^\top \oplus \bm{H}\cdot\mathds{1}_i^\top \oplus \bm{H}\cdot\mathds{1}_j^\top \oplus \bm{H}\cdot\mathds{1}_k^\top.
\label{eq:three-bits-constraint}
\end{equation}

Equations (\ref{eq:one-bit-constraint})-(\ref{eq:three-bits-constraint}) are the core of the proposed 
VLSI architecture. By combining several different one bit-flip noise sequence syndromes, 
it is possible to compute all the queries corresponding to several bit-flips. 
In the following, we denote by $\bm{s}_i$ the syndrome corresponding to the one-bit-flip error pattern at location $i$: 
$\bm{s}_i = \bm{H}\cdot\mathds{1}_i^\top$, which also corresponds to the $i^{\text{th}}$ column of the parity check matrix.

\subsection{Principle, Details and Scheduling}\label{sec:arch_principle}
The scheduling of the proposed architecture comprises four fundamental decoding steps. In the first one, 
the syndrome of the received word is computed ($\bm{H}\cdot\bm{r}^\top$). In the second step, all the error patterns with a 
Hamming weight of 1 are independently combined with the syndrome of the received word. In the third and fourth steps, 
Hamming weights of 2 and 3 are considered, respectively.
During the iterations of any of the described steps, when (\ref{eq:constraint}) results in a zero, the corresponding estimated word is the  
output and the procedure is terminated.
To efficiently generate the different error patterns, the proposed architecture is based on what we call \emph{dials}.

\begin{figure}[!t]
\centering
    \includegraphics[height=1.75in]{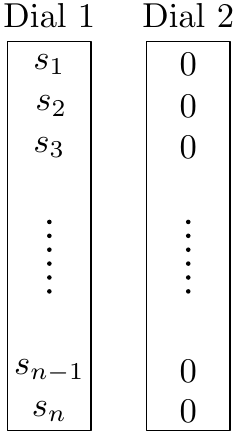}
    \caption{\label{fig:dial_1bf} Content of the dials for checking the one-bit-flip error patterns.}
\end{figure}

A dial is a $n\times (n-k)\text{-bit}$ register file which stores all the $n$ syndromes associated with the one-bit-flip 
error patterns ($\bm{s}_i$). The dial has the ability to shift its content in a cyclic manner at each time step; 
\emph{i.e.} when the content of the second row is shifted to the first row, the content of the first row is shifted to
the last row. Moreover, during a cyclic shift, the content of the last row may be replaced by the $(n-k)\text{-bit}$ wide 
null vector. This operation is called \emph{shift-up}. After a shift-up operation has taken place, the following cyclic shifts 
exclude rows containing null vectors. Note that a dial works in conjunction with an \emph{index dial}, a 
$n\times \log_2n\text{-bit}$ cyclic shift register file, which performs the same operations 
(cyclic shift or shift-up) to keep track of the indices ($i$ in $\bm{s}_i$). As explained later, only 2 dials are used
in the proposed architecture.

For checking the one-bit-flip error patterns, the content of the dials is depicted in Fig. \ref{fig:dial_1bf}. By 
combining each row of the dials with the syndrome of the received vector, we can compute (\ref{eq:one-bit-constraint}) in one time step. 

\begin{figure}[!t]
    \captionsetup[subfloat]{farskip=0pt}
    \centering
    \subfloat[First time step.]
    {
        \includegraphics[height=1.75in, page=1]{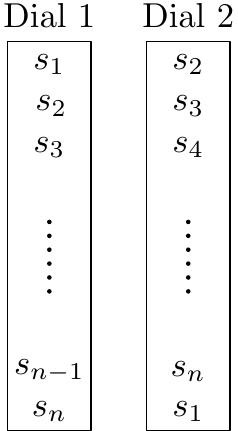}
        \label{fig:dial_2bf:a}
    }\hfil
    \subfloat[Second time step.]
    {
        \includegraphics[height=1.75in, page=2]{figs/dial_2bf.pdf}
        \label{fig:dial_2bf:b}
    }
    \caption{\label{fig:dial_2bf} Content of the dials for checking the two-bit-flips error patterns at different time steps.}
\end{figure}

Fig. \ref{fig:dial_2bf}(a) shows the content of the dials at the first time step when checking for the two-bit-flips 
error patterns: the content of the dial 2 is the image of dial 1 cyclically shifted by one. By combining each row of the dials 
with the syndrome of the received vector, we can compute $n$ two-bit-flips error patterns in one time step. 
At the next time step, the content of the dial 2 is cyclically shifted by one as shown in Fig. \ref{fig:dial_2bf}(b).
Observing that $\mathds{1}_{i,j} = \mathds{1}_{j,i}$, all the $\binom{n}{2}$ two-bit-flips error patterns are 
tested for code membership after a total of $\left\lfloor\frac{n}{2}\right\rfloor-1$ cyclic shifts from the original setting (Fig. \ref{fig:dial_2bf}(a)). 
Hence, a total of $\left\lfloor\frac{n}{2}\right\rfloor$ time steps are required to compute (\ref{eq:two-bits-constraint}).
Note that to keep track of the indexes, whenever a dial is rotated, its corresponding index shift register (index
dial) is also rotated.

\begin{figure}[!t]
    \centering
    \captionsetup[subfloat]{farskip=0pt}
    \subfloat[First time step.]
    {
        \includegraphics[height=1.75in, page=1]{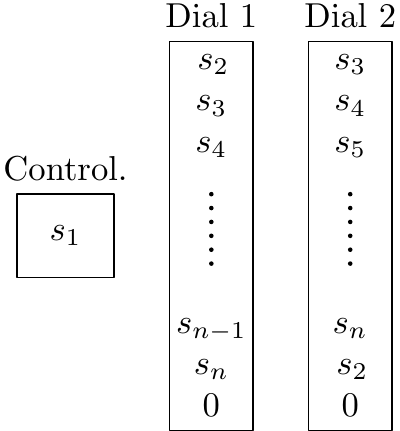}
        \label{fig:dial_3bf:a}
    }\hfil
    \subfloat[Second time step.]
    {
        \includegraphics[height=1.75in, page=2]{figs/dial_3bf.pdf}
        \label{fig:dial_3bf:b}
    }

    \captionsetup[subfloat]{farskip=10pt}
    \subfloat[Time step $\left\lfloor\frac{n-1}{2}\right\rfloor+1$.]
    {
        \includegraphics[height=1.75in, page=3]{figs/dial_3bf.pdf}
        \label{fig:dial_3bf:c}
    }\hfil
    \subfloat[Time step $\left\lfloor\frac{n-1}{2}\right\rfloor+2$.]
    {
        \includegraphics[height=1.75in, page=4]{figs/dial_3bf.pdf}
        \label{fig:dial_3bf:d}
    }
    \caption{\label{fig:dial_3bf} Content of the dials and syndrome outputted by the controller for checking the 
    three-bit-flips error patterns at different time steps.}
\end{figure}

Regarding the three-bit-flips error patterns, we show that only two dials can be used. Indeed, if three dials are considered, 
the scheduling and the associated hardware become more complex to avoid error pattern duplications.
Instead, a controller is used in conjunction with the dials to generate the test 
patterns. The controller takes care of the first bit-flip, while the dials are responsible for considering the two other 
bit-flips. Fig. \ref{fig:dial_3bf}(a) shows the content of the dials and the syndrome output by the controller to 
generate $n-1$ three-bit-flips error patterns at time step 1. To do so, the dial 1 is shifted-up by 1, while the
dial 2 is shifted-up by 1 and cyclically shifted by 1 at the initialization. In the next time step, the dial 2 is cyclically shifted
by 1 to generate the next $n-1$ three bit-flip noise sequences as shown in Fig. \ref{fig:dial_3bf}(b).
After $\left\lfloor\frac{n-1}{2}\right\rfloor$ time steps all the 
$\binom{n-1}{2}$ three-bit-flips error patterns with $\bm{s}_1$ are generated.
In the next time step, the controller outputs $\bm{s}_2$ while 
the dial 1 is shifted-up by 1 and the dial 2 is reset, shifted-up by 2 and cyclically shifted by 1. This generates $n-2$ 
three-bit-flips error patterns, as shown in Fig. \ref{fig:dial_3bf}(c). In the next time step, the dial 2 is cyclically 
shifted by 1, allowing to generate the next $n-2$ three-bit-flips error patterns as shown in Fig. \ref{fig:dial_3bf}(d).
Hence, $\left\lfloor\frac{n-2}{2}\right\rfloor$ time steps are 
used to generate all the $\binom{n-2}{2}$ three-bit-flips error patterns with $\bm{s}_2$ set and $\bm{s}_1$ excluded.
Similarly, this process is repeated until $\bm{s}_{n-2}$ is outputted by the controller, where only one three-bit-flips
error pattern is generated: $\bm{H}\cdot\bm{r}^\top \oplus \bm{H}\cdot\bm{s}_{n-2}^\top \oplus \bm{H}\cdot\bm{s}_{n-1}^\top \oplus \bm{H}\cdot\bm{s}_{n}^\top$.
Finally, checking all the three-bit-flips error patterns requires $\sum_{i=2}^{n-1} \left\lfloor\frac{i}{2}\right\rfloor$
time steps.

In summary, the number of required time steps to check all the error patterns with Hamming weights of 3 or less is 
given by: 
\begin{equation}
\label{eq:nb_steps}
2+\sum_{i=2}^{n} \left\lfloor\frac{i}{2}\right\rfloor.
\end{equation}
Using some mathematical manipulation, the ratio between (\ref{eq:queries}) and (\ref{eq:nb_steps}) -- that expresses the 
parallelization factor -- can be approximated by $\frac{2*n}{3}$. Thus, the longer the code, the higher the savings compared with a conventional
and serial approach.


\begin{figure*}[!htb]
    \centering
    \includegraphics[width=1\textwidth]{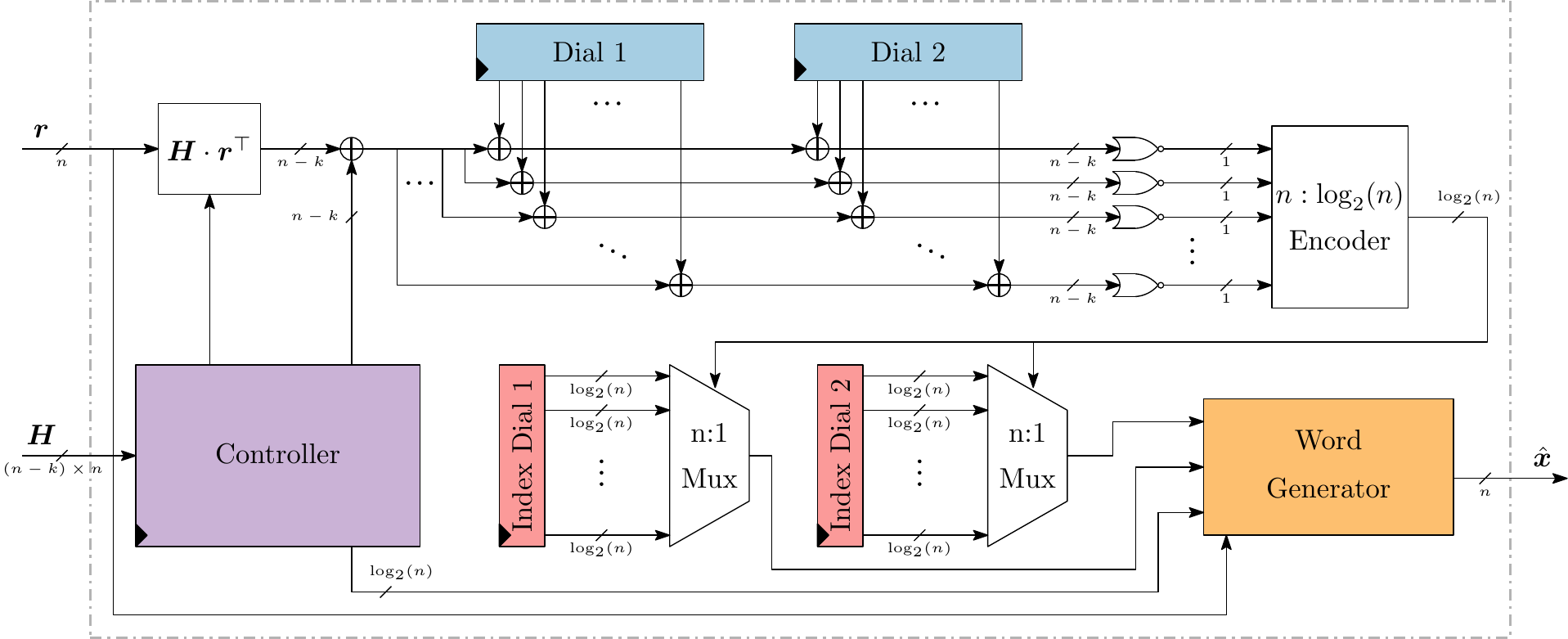}
    \caption{\label{fig:final_arch} Proposed architecture for GRANDAB ($\text{AB}=3$).} 
\end{figure*}

The proposed hardware architecture for the GRANDAB algorithm with $\text{AB}=3$ is shown in Fig. \ref{fig:final_arch}. 
Its input is the hard decision vector $\bm{r}$ of length $n$ and its output is the estimated word $\hat{\bm{u}}$, 
padded with zeros to match the length of $n$. For the sake of clarity, the control and clock signals are omitted in the 
Figure. At any time, to support any code given the length and rate constraints, an $\bm{H}$ matrix can be loaded. The data path consists essentially of the 
interconnection through $2\times n +1$ $(n-k)$-bit-wide XOR gates of the dials, the syndrome of the received word, and the syndrome 
provided by the controller ($2\times n$ for the dials and $1$ for the controller), as described in the previous paragraphs.
Each of the $n$ test syndromes is NOR-reduced, to feed an $n$-to-$\log_2n$ priority encoder. The output of each NOR-reduce is 1 
if and only if all the bits of the syndrome computed by (\ref{eq:constraint}) are 0. The output of the priority 
encoder controls two multiplexers, used to forward the indices associated with the valid syndrome to the word 
generator. Finally, the word generator combines the hard decision vector $\bm{r}$ and the three indices to produce the 
estimated codeword, which is translated into the estimated word and outputted.

\section{Implementation Results}
The proposed architecture has been implemented in Verilog HDL and synthesized using the Synopsys Design 
Compiler version P-2019.03 with TSMC 65nm CMOS technology. The design has been verified using test benches generated via the bit-true C model 
of the proposed hardware. 

\begin{figure}[!t]
\centering
  \begin{tikzpicture}
    \begin{groupplot}[group style={group name=lat_tp, group size= 2 by 1, horizontal sep=10pt, vertical sep=10pt}, 
                      footnotesize,
                      height=.6\columnwidth,  width=.55\columnwidth,
                      xlabel=SNR (dB),
                      xmin=6, xmax=11, xtick={4,5,...,12},
                      ymode=log,
                      tick align=inside, 
                      grid=both, grid style={gray!30},
             ]

      \nextgroupplot[ylabel= Latency (cycles), ytick pos=left, y label style={at={(axis description cs:-0.15,.5)},anchor=south},ymin=1, ymax = 2e3]
        \addplot[mark=diamond , Paired-7, semithick]  table[x=SNR, y=Lat120] {data/crc/throughput.txt};\label{gp:plot1}
        \addplot[mark=triangle, Paired-5, semithick]  table[x=SNR, y=Lat112] {data/crc/throughput.txt};\label{gp:plot2}
        \addplot[mark=square  , Paired-3, semithick]  table[x=SNR, y=Lat104] {data/crc/throughput.txt};\label{gp:plot3}
        \addplot[mark=o       , Paired-1, semithick]  table[x=SNR, y=Lat96]  {data/crc/throughput.txt};\label{gp:plot4}
        \coordinate (top) at (rel axis cs:0,1);

      \nextgroupplot[ylabel=Info. Throughput (Mbps), ytick pos=right,y label style={at={(axis description cs:1.33,.5)},anchor=south}, ymax = 1e5]
        \addplot[mark=diamond , Paired-7, semithick]  table[x=SNR, y=TP120] {data/crc/throughput.txt};
        \addplot[mark=triangle, Paired-5, semithick]  table[x=SNR, y=TP112] {data/crc/throughput.txt};
        \addplot[mark=square  , Paired-3, semithick]  table[x=SNR, y=TP104] {data/crc/throughput.txt};
        \addplot[mark=o       , Paired-1, semithick]  table[x=SNR, y=TP96]  {data/crc/throughput.txt};

        \coordinate (bot) at (rel axis cs:1,0);
    \end{groupplot}
    \node[below = 1cm of lat_tp c1r1.south] {(a) : Latency};
    \node[below = 1cm of lat_tp c2r1.south] {(b) : Info. Throughput};
    \path (top|-current bounding box.north) -- coordinate(legendpos) (bot|-current bounding box.north);
    \matrix[
        matrix of nodes,
        anchor=south,
        draw,
        inner sep=0.2em,
        draw
      ]at(legendpos)
      {
        \ref{gp:plot1}& \footnotesize CRC(120,128) &[5pt]
        \ref{gp:plot2}& \footnotesize CRC(112,128) \\
        \ref{gp:plot3}& \footnotesize CRC(112,128) &[5pt]
        \ref{gp:plot4}& \footnotesize CRC(128,\hphantom{1}96)\\};
  \end{tikzpicture}
  \caption{\label{fig:lat_tp}Average latency and average information throughput of the proposed hardware architecture, using the same coding schemes as in Fig. \ref{fig:fer_crc}.}
\end{figure}
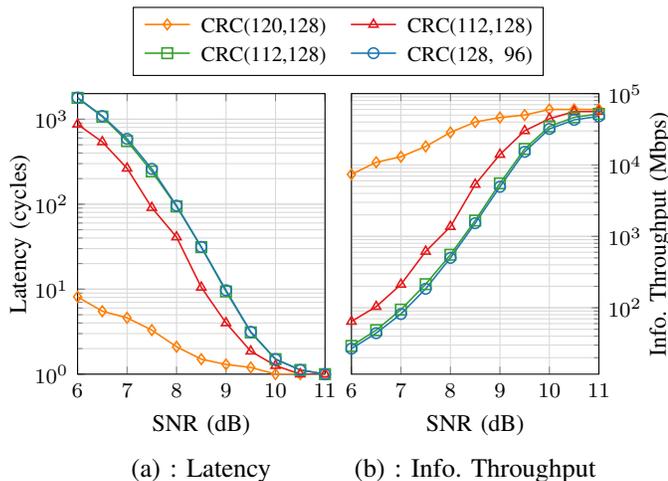

Table \ref{table:synth_128} presents the synthesis results for the proposed decoder with $n=128$, $\text{AB}=3$ and a 
code rate between $0.75$ and $1$. Thus, the length of the syndromes is constrained to the interval $\llbracket 0 \isep 32 \rrbracket$.
The implementation can support a maximum frequency of $500~\text{MHz}$. No pipelining strategy is used, therefore one clock 
cycle corresponds to one time step. Using (\ref{eq:nb_steps}), $\numprint{4098}$ cycles are required in the worst-case (W.C.)
for decoding a 128-length code. Recall that GRANDAB ($\text{AB} = 3$) requires a total number of $\numprint{349632}$ 
queries for decoding any code of length $128$. Hence, our proposed VLSI architecture demonstrates only a fraction ($1.2\%$) of 
the total number of performed queries as latency.
With a frequency of $500~\text{MHz}$, the proposed architecture results in a worst-case information throughput (W.C. T/P) of 
$11.71$ to $14.64$ Mbps for the CRC codes considered in Section \ref{sec:grandab}. However, the average latency is much shorter than the worst-case latency, 
especially in the mid-to-high SNR region. Using the bit-true model, the average latency is computed after considering at least 100
frames in error for each SNR points. Fig. \ref{fig:lat_tp}(a) depicts the average latency for the considered codes. 
Irrespective of the code rate, we can see that the average latency reduces when the channel condition becomes better, 
up to the point where the average latency reach only 1 cycle per decoded codeword. 
The counterpart of the latency, the throughput, is given in Fig. \ref{fig:lat_tp}(b). Observe that
the information throughput grows with the SNR up to reaching the values of $48$ Gbps to $60$ Gbps, according to the code rate.
In addition, considering an FER of $10^{-4}$, average information throughputs of $9$ Gbps, $9$ Gbps, $56$ Gbps and $60$ Gbps
are obtained for the information lengths of $96$, $104$, $112$, and $120$, respectively. 

\begin{table}[!t]
\centering
\caption{TSMC 65 nm CMOS Implementation Results for GRANDAB ($\text{AB} = 3$) and $n=128$.}
\label{table:synth_128}
\begin{tabular}{@{}lrr@{}}
\toprule
Technology (nm)                  &             & $65   $ \\ 
Supply (V)                       &             & $0.9  $ \\
Max. Freq (MHz)                  &             & $500  $ \\
Area ($\text{mm}^2$)             &             & $0.25 $ \\
W.C. Latency (cycles)            &             & $4098 $ \\
\multirow{4}{*}{W.C. T/P (Mbps)} & $(128,\hphantom{1}96)$ & $11.71$ \\
                                 & $(128,104)$ & $12.68$ \\ 
                                 & $(128,112)$ & $13.66$ \\
                                 & $(128,120)$ & $14.64$ \\
\bottomrule
\end{tabular}
\end{table}

\begin{table}[!t]
\centering
\caption{\label{table:synth_79}TSMC 65 nm CMOS Implementation Comparison for GRANDAB ($\text{AB} = 2$) and $n=79$.}
\begin{adjustbox}{max width=\columnwidth}
\begin{tabular}{@{}llrr@{}}
\toprule
                              &      & GRANDAB ($\text{AB} = 2$)    & $(79,64)$ BCH decoder \cite{Choi19}     \\
                              \cmidrule(lr){3-3}\cmidrule(l){4-4}
Technology (nm)               &      & 65                           & 65                                    \\
Supply (V)                    &      & 1.1                          & 1.2                                   \\
Frequency (GHz)               &      & 1                            & N/A                                   \\
Area ($\mu\text{m}^2$)        &      & \numprint{126733}            & \numprint{3264}                       \\
\multirow{3}{*}{Latency (ns)} & min. & 1                            & 1.1                                   \\
                              & avg. & 1.09                         & 1.1                                   \\
                              & max. & 41                           & 3                                     \\
Code compatible               &      & Yes                          & No                                    \\
Rate compatible               &      & Yes                          & No                                    \\
\bottomrule
\end{tabular}
\end{adjustbox}
\end{table}

To the best of our knowledge, there is no hardware implementation of a hard detection decoder in the literature that 
achieves the same code flexibility as our proposed architecture. Thus, performing a fair comparison is difficult. 
However, we propose to compare it with a state-of-the-art hard decision algebraic decoder. 
Recently, a high throughput VLSI architecture based on the PGZ algorithm for a $(79,64)$ BCH code decoder has been proposed \cite{Choi19}. 
Since up to 2 errors can be corrected with this decoder, 
GRANDAB with $\text{AB}=2$ is enough to achieve the same decoding
performance. Therefore, we re-synthesized our architecture by limiting $n$ to $79$ and by
setting $\text{AB}=2$. Hence, a total of 
$1+1+\left\lfloor\frac{79}{2}\right\rfloor = 41$ time steps are required to decode any code of length $79$ with at
most 2 errors. 

Table \ref{table:synth_79} compares the implementation results of the GRANDAB ($\text{AB} = 2$) decoder and the BCH decoder
in \cite{Choi19}. The proposed decoder is $41\times$ larger and has $13.6\times$ higher worst-case latency. On the other 
hand, the average latency of the two decoders are equivalent at an SNR of $10\text{dB}$. At higher SNRs, the proposed 
decoder exhibits a slightly better minimum latency and achieves an information throughput of $64$ Gbps, while the BCH 
decoder is limited to 58 Gbps. Finally, while \cite{Choi19} can only decode the $(79,64)$ BCH code, the proposed GRANDAB 
 ($\text{AB} = 2$) can decode any code with $n=79$ and $R\geq0.75$.

\section{Conclusion}
In this paper, we proposed the first hardware architecture for the GRANDAB algorithm. The decoding algorithm has 
the uncommon property of being able to decode any linear code. By using linear algebra basics, we were able to decompose
the computations of the GRAND algorithm to improve the inherent parallelism. By doing so, the proposed hardware architecture can accomplish 
$\numprint{349632}$ queries in $\numprint{4098}$ time steps. ASIC synthesis results showed that an average information 
throughput of at least $9$ Gbps can be achieved with a block length of $128$ when a FER of $10^{-4}$ is targeted. 
Moreover, the average throughput increases when the channel conditions become better. Hence, 
the average coded throughput for the same parameters can reach up to $64$ Gbps.
Finally, the architecture can achieve the same average throughput as a BCH decoder tailored for a $(79,64)$ code. 
The proposed architecture paves the way for future implementation of the GRAND algorithm that can consider soft information as their inputs.
\balance
\bibliographystyle{IEEEtran}
\bibliography{IEEEabrv, GRANDv2}
\end{document}